\documentclass[sigconf]{acmart}
\setcopyright{rightsretained}

\usepackage[normalem]{ulem}
 \useunder{\uline}{\ul}{}

\usepackage{subfig}
\usepackage{subfloat}

\begin{document}
\title{A Comparative Case Study of HTTP Adaptive Streaming Algorithms in Mobile Networks}

\author{Theodoros Karagkioules$^{*\dag}$, Cyril Concolato$^{\dag}$}
\orcid{1234-5678-9012}

\affiliation{%
	\institution{$^{\dag}$LTCI, T\'el\'ecom ParisTech, Universit\'e Paris-Saclay}
}
\email{{karagkioules, cyril.concolato}@telecom-paristech.fr}
\author{Dimitrios Tsilimantos$^{*}$, Stefan Valentin$^{*}$}
\orcid{1234-5678-9012}
\affiliation{%
	\institution{$^{*}$Mathematical and Algorithmic Sciences Lab, France Research Center,  Huawei Technologies France SASU}
}
\email{{dimitrios.tsilimantos, stefan.valentin}@huawei.com}

\renewcommand{\shortauthors}{ T. Karagkioules et al.}
\newcommand{\TODO}[1]{{\color{blue} \bfseries #1}}

\begin{abstract}
HTTP Adaptive Streaming (HAS) techniques are now the dominant solution for video delivery in mobile networks. Over the past few years, several HAS algorithms have been introduced in order to improve user quality-of-experience (QoE) by bit-rate adaptation. Their difference is mainly the required input information, ranging from network characteristics to application-layer parameters such as the playback buffer. Interestingly, despite the recent outburst in scientific papers on the topic, a comprehensive comparative study of the main algorithm classes is still missing. In this paper we provide such comparison by evaluating the performance of the state-of-the-art HAS algorithms per class, based on data from field measurements. We provide a systematic study of the main QoE factors and the impact of the target buffer level. We conclude that this target buffer level is a critical classifier for the studied HAS algorithms. While buffer-based algorithms show superior QoE in most of the cases, their performance may differ at the low target buffer levels of live streaming services. Overall, we believe that our findings provide valuable insight for the design and choice of HAS algorithms according to networks conditions and service requirements. 
\end{abstract}


\copyrightyear{2017} 
\acmYear{2017} 
\setcopyright{acmcopyright}
\acmConference{NOSSDAV'17}{June 20-23, 2017}{Taipei, Taiwan}\acmPrice{15.00}\acmDOI{http://dx.doi.org/10.1145/3083165.3083170}
\acmISBN{978-1-4503-5003-7/17/06}

\maketitle
\section{Introduction}
\label{introsection}

Mobile video accounted for 60\% of the global mobile data traffic in 2016 and this percentage is projected to further increase and reach a striking 78\% by 2021 \cite{cisco2017}. Most of this traffic is video-on-demand (VoD) streaming via  HTTP Adaptive Streaming (HAS) \cite{Sandvine}, which undoubtedly becomes fast an integral part of the mobile client's life. In order to keep pace with this explosion of video traffic, significant progress has been made to the development and design of adaptive streaming solutions and standards. For instance, dynamic adaptive streaming over HTTP (MPEG-DASH) is an international standard that uses the existing HTTP web server infrastructure and has become very popular in the last years \cite{Sodagar}.

The main characteristic of HAS solutions that led to their vast deployment in the market is their ability to adjust the play-out quality during the video session. Their target is to deliver the highest possible quality-of-experience (QoE) given the dynamic nature of the wireless channel conditions and the presence of diverse bottlenecks in the video delivery system. In order to achieve that, video files are encoded in various quality representations which are then stored in a web server. Each representation is subdivided in smaller files called segments, usually of constant duration and variable size due to the commonly adopted variable bit-rate (VBR) encoding. After obtaining a manifest file with all the necessary video information, the client sequentially requests and downloads each segment in the quality indicated by the algorithm of the deployed HAS algorithm.

User perceived QoE plays a critical role for the assessment of the various HAS solutions, since it is directly connected to the user engagement and thus, the revenue of content providers. In particular, video stalls and frequent video bit-rate switching are dominating QoE factors for mobile HAS \cite{HAS_QoE}. Avoiding stalls due to the depletion of the client's play-back buffer, while at the same time minimizing the frequency of adaptation and providing high average quality, is a very challenging task, especially at high network load or at poor wireless coverage. The inherent trade-offs between the key video QoE metrics (i.e quality, stability and smoothness) makes this attempt even more difficult.

In order to address these problems, several HAS algorithms have been proposed recently, which can be classified into three main categories with respect to the required input information. Firstly, throughput-based algorithms, such as PANDA \cite{Panda} or Festive \cite{Festive}, rely their decision on the observed TCP throughout, which requires a sufficient number of probes to obtain reliable measurements. Secondly, time-based algorithms such as ABMA+ \cite{abmaplus} rely on the same principle of probing, but this time to estimate the download time of each segment. Lastly, buffer-based algorithms, such as BBA \cite{Netflix} and BOLA \cite{BOLA}, observe and react to the level of the client's playback buffer. Despite several recent research efforts and proposals, there still appears to be a lack of consensus and an ongoing debate regarding the merits of the above classes of algorithms.

In this paper we try to shed some light to that debate. Specifically, we investigate throughput-based, time-based and buffer-based adaptation algorithms using a set of commonly studied traces of mobile throughput measurements. Since buffer-based adaptation algorithms have not been included before in similar comparisons, the main scope of this work is to provide some insight on the merits of each class of algorithms. Our comparison is based on the implementation of the algorithms in a single simulation framework, which uses throughput information from publicly available network profiles. Furthermore, due to the increasing popularity of live streaming services, which have small buffer sizes due to the strict real-time delay requirements, we also examine the impact of the buffer size on the performance of the HAS algorithms. Therefore another key contribution of this work is the consideration of two typical maximum occupancy buffer levels to investigate the performance for both live and VoD streaming scenarios.

Several similar comparisons can be found  in  literature. In particular, in \cite{6774590} the authors investigate typical adaptation methods in the context of live video streaming. In \cite{DBLP:journals/corr/TimmererMR16} the authors make both subjective and objective studies of various throughput-based adaptive streaming algorithms, but no other class of algorithms is included in this work. \cite{Muller:2012:EDA:2151677.2151686}  is a similar experimental evaluation of HAS algorithms on mobile vehicular networks. We have found that a study that focuses on the categorization of the algorithms according to their input dynamics, along with a performance evaluation on mobile vehicular networks that considers live streaming as well, is missing from the literature. To this end, our work presents a comparison of the latest HAS algorithms, per class, which includes buffer-based and time-based solutions for the first time.

The remainder of the paper is organized as follows. In Section \ref{HASpoliciessection} we briefly describe the main principles and properties of the selected state-of-the-art HAS algorithms. In Section \ref{experimentalframeworksection} we describe the validation process followed to obtain our comparison results, including our implementation parameters and simulation factors. Then, in Section \ref{resultssection} we present our results on the performance of the HAS algorithms. 
 We conclude with our remarks in Section \ref{conclusionssection}.

\section{Adaptive streaming algorithms}
\label{HASpoliciessection}
In  this section we briefly present the five state-of-the-art HAS algorithms that were studied, implemented and compared. We have chosen the most representative algorithms, per class, as they have been used in literature for other comparisons. 
\subsection{Throughput-based adaptation}
A very common TCP throughput-based adaptation scheme \cite{Festive,Panda} is based on a four-step adaptation model, where initially the available network bandwidth is estimated and then smoothed using noise-filters to avoid estimation errors due to throughput variation. Then, the video bit-rate is indicated based on the discretized output of the smoothing step. The next segment request is scheduled once the inter-request time is estimated.

\emph{Conventional} is a simple adaptation algorithm, based on the four step model, which equates the current available bandwidth with the TCP throughput, as it is measured during the previous segment download. Then, the proposed video bit-rate is yielded by applying an exponential weighted moving average (EWMA) filter and a dead-zone quantizer. The algorithm determines the inter-request time of the next segment using a bi-modal scheduler, by which the next segment request is scheduled either with a constant delay when the buffer is full or immediately otherwise.

\emph{PANDA} \cite{Panda} is an advanced variation of the four-step model, yet with two distinct modifications. In the estimation step this algorithm uses a more proactive probing mechanism, that is designed to minimize video bit-rate oscillations. The second modification is at the scheduling step, where a more sophisticated scheduler is considered that drives the buffer level towards  the maximum buffer occupancy level $B_{max}$. At the same time the inter-request time is matched to the necessary time needed to complete the download based on the smoothed estimated value of the available bandwidth.
\subsection{Buffer-based adaptation} 
\emph{BBA} is a very well known  buffer-based adaptation algorithm. In  \cite{Netflix}, the authors introduce a segment map based on the average size of the segments for every representation. The map is defined by two thresholds: i) an upper threshold that drives the policy to select the maximum quality available ($R_{max}$), once the instantaneous buffer occupancy surpasses it and ii) a lower threshold that dictates the lowest  available quality ($R_{min}$), if the buffer is lower than that threshold. In the buffer region between these thresholds the policy may use any non decreasing function to select the quality of the next requested segment.

\emph{BOLA} \cite{BOLA} is a buffer-based adaptation algorithm that uses Lyapunov optimization in order to indicate the video bit-rate of each segment. Practically, the algorithm is designed to maximize a  joint utility function that rewards an increase in the average quality and penalizes potential re-buffering occurrences. More specifically, a variation called \emph{BOLA-O}, mitigates video bit-rate oscillations by introducing a form of bit-rate capping when switching to higher bit-rates.
\subsection{Time-based adaptation}
 Download time is considered as a higher level parameter than throughput, thus, in this study, time-based adaptation is treated as a separate class of algorithms. \emph{ABMA+} \cite{abmaplus}  is an adaptation and buffer management algorithm, which selects the video representation based on the predicted probability of video stalling. The algorithm continuously estimates the segment download time and uses a pre-computed play-out buffer map to select the maximum video representation, which guarantees smooth content play-out. The segment download time estimation is based on the same probing mechanism as the throughput-based method, but \emph{ABMA+} takes into account VBR aspects as well.


\section{Experimental framework}
\label {experimentalframeworksection}

\subsection{Selected network data-sets}

 The performance of HAS algorithms is highly correlated with the network conditions during the streaming. As opposed to fixed networks, mobile networks are characterized by their intense throughput variation. Additionally, due to diverse coverage quality  there may appear areas with prolonged low  bandwidth, which will result to throughput outages and therefore an increased probability of a video stall. In order to avoid stalls, an HAS algorithm needs, at the very least, a network profile that offers a mean throughput at least higher than the lowest available representation stored at the server. Otherwise, the buffer may be completely depleted, leading to a stall event.
\begin{figure}[t]
\includegraphics[width=1\linewidth,clip]{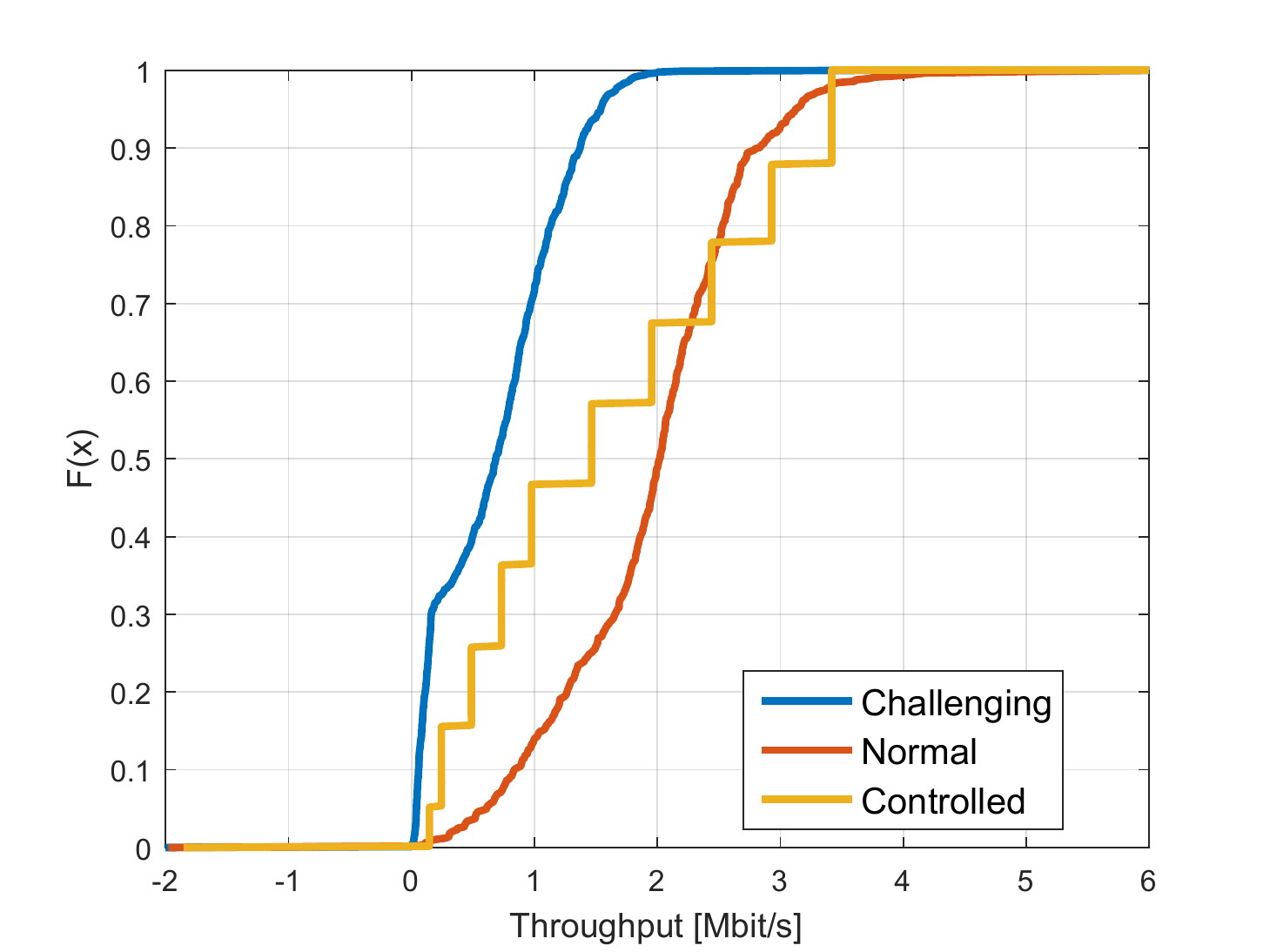} \vspace{-15pt}
\caption{Empirical CDF of throughput for the selected network profiles.} 
\label{fig:TotaleCDF} \vspace{-10pt}
\end{figure}
\begin{figure}[t]
\includegraphics[width=1\linewidth,clip]{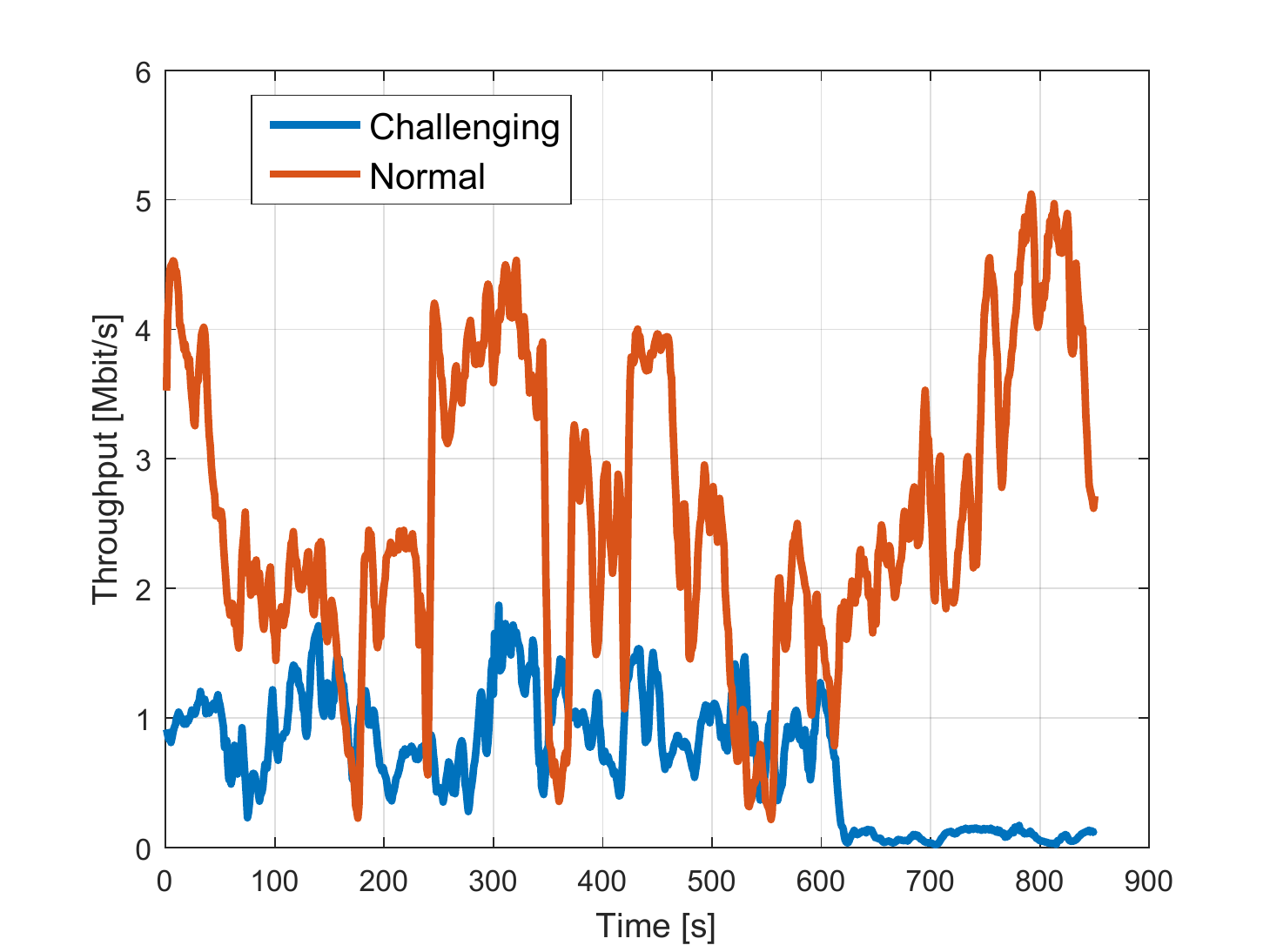} \vspace{-15pt}
\caption{Characteristic example of selected real-field network profiles}
\label{fig:Profiles}
\end{figure}

In order to obtain insightful results for our comparison, we  chose two diverse  throughput profiles for our simulations, that are representative of a normal and a challenging network profile in vehicular environments. These profiles correspond to direct throughput measurements from a bus and an underground metro respectively \cite{oslotraces}. We investigate mobile networks as they are capable of stressing the adaptation methods to highly challenging conditions, as opposed to fixed networks. In particular, we preferred the use of 3G traces as LTE, although more contemporary, offers higher throughput which is not always experienced by the user.  Additionally  we use an artificial profile, which offers controlled network conditions in order to validate our implementations. In Fig. \ref{fig:TotaleCDF} we can see the cumulative distribution function (CDF) of all studied network profiles.

The \emph{controlled} profile corresponds to a High-Low-High network profile inspired from \cite{Forum2014}. This profile is shown in Fig. \ref{fig:QS} and it is characterized by the distinct and controlled increases and decreases of the total throughput every 30 s. The \emph{normal} profile is illustrated in  Fig. \ref{fig:Profiles}. It is characterized by significant bandwidth variation, which is expected from real networks. In general a high throughput is sustained and no significant outages appear. This profile was  chosen as representative of a vehicular terrestrial network profile and corresponds to the "bus" data-set as described in \cite{oslotraces}, consisting of 5 traces, after excluding those that showed long outages. The \emph{challenging} profile corresponds to measurements made on an underground metro and consists of a selection of 7 traces which, in general, show a low throughput throughout the route and there is a long outage period when the metro enters a tunnel towards the end of the trace, as depicted in Fig. \ref{fig:Profiles}. This profile allows us to stress the selected HAS algorithms and test their performance under difficult and extreme conditions. We expect to see an increased re-buffering frequency with this "underground" trace-set.

\begin{table}[!t]
	\centering
	\caption{Video representations}
	\label{vireso}
	\resizebox{1.0\linewidth}{!}{\begin{tabular}{|c|c|c|c|c|c|}
			\hline
			\textbf{\begin{tabular}[c]{@{}c@{}}Representation\\ index\end{tabular}}& \textbf{Resolution} & \textbf{\begin{tabular}[c]{@{}c@{}}BBB\\Max encoding\\ rate (kbps)\end{tabular}}  & \textbf{\begin{tabular}[c]{@{}c@{}}TSA\\Max encoding\\ rate (kbps)\end{tabular}}  & \textbf{\begin{tabular}[c]{@{}c@{}}RBPS\\Max encoding\\ rate (kbps)\end{tabular}} & \textbf{\begin{tabular}[c]{@{}c@{}}CDF\\Quantiles\end{tabular}} \\ \hline
			1                      & 320$\times$240      & 129                                                                        & 128                                                                           & 149                       &        0.01                               \\ \hline
			2                      & 480$\times$360      & 378                                                                       & 330                                                                            & 395                       &        0.05                                      \\ \hline
			3                      & 854$\times$480    & 578                                                                            & 754                                                                         & 700                        &       0.1                         \\ \hline
			4                      & 1280$\times$720     & 985                                                                   & 1331                                                                           & 1536                         &      0.25                \\ \hline
			5                      & 1280$\times$720   & 1536                                                                          & 2048                                                                          & 2048                    &     0.5                            \\ \hline
			6                      & 1920$\times$1080    & 2353                                                                          & 2764                                                                        &2560                &         0.75                                                       \\ \hline
			7                      & 1920$\times$1080    & 2969                                                                        &  3481                                                                       & 3072                 &         0.95                                                  \\ \hline
		\end{tabular}} 
	\end{table} 

\subsection{Streaming content}
As streaming content, we have chosen 3 representative open movies commonly used  for testing video codecs and streaming protocols and recommended in the measurement guidelines of the DASH Industry Forum \cite{Forum2014}. The first movie is \emph{ Big Buck Bunny (BBB)}, a high motion computer animated movie of 9:56 min duration. The second is \emph{The Swiss Account (TSA)},  which is a sport documentary with regular motion scenes and a duration  of 57:34 min. The third is \emph{Red Bull Play Street (RBPS)}, which is a sport show with high motion scenes and of 1:37 hours duration.
For all movies we used the video encodings of \cite{ITEC} in order to obtain the representation levels $R_i$, where $i=\{1,2,\dots,N\}$. 
We selected a total of $N=7$ video bit-rate levels, based on the quantiles of the CDF distribution of the normal network profile (Table. \ref{vireso}), with a segment duration of $4$ s. The particular selection of the representations was made in order to ensure that the minimum representation level is sustainable for 99.9\%  of the normal profile. Of course this would lead to a very small probability of re-buffering for that case, yet it serves as a good basis for the comparison with the challenging profile. Additionally, we chose a high number of distinct representations in order to make the transitions smoother between quality switches. QoE studies \cite{HAS_QoE} suggest that adaptation amplitude is the dominant adaptation factor, which means that finer granularity switching may compensate for higher switching frequency. One movie was used per trace, chosen at random to ensure unbiased statistics, and it was repeated if the trace duration was larger than its duration.

\subsection{Client model and metrics }
The client model consists of the maximum buffer level $B_{max}$ and the selected HAS algorithm that the player may deploy during a video streaming session. We ran our experiments over 12 mobile traces (7 normal network traces and 5 challenging network traces). Furthermore, we investigated the maximum buffer occupancy factor $B_{max}$, since various applications (live, VoD, short clips or long movies)  may target different maximum buffer occupancy levels. In particular we repeated the experiment for a small $B_{max}=16$ s (4 segments), which simulates a live streaming service and for a larger $B_{max}=92$ s (23 segments), to simulate the case of VoD. These studied values were selected based on measurements of the maximum buffer level of a popular streaming service, which offers both Live and VoD streaming. We assert that our results hold for any buffer value larger than 4 segments, but leave the full study of the impact of the buffer level to future work. Also two important parameters that may affect the QoE of the user are the initial buffering (i.e the amount of segments that need to be downloaded in the buffer before play-out can start initially) and the re-buffering threshold (i.e the amount of segments that need to be downloaded in the buffer before play-out can resume after a stall event). These parameters were both set equal to $\omega=2$ segments for our experiments, as indicated in most of the implementation guidelines of the proposed algorithms. 

Although a unified framework for measuring QoE is missing from the literature, several related works \cite{HAS_QoE, 6178830} suggest that adaptability, instability and un-smoothness of streaming are the most important elements for quantifying QoE in an objective manner.  Inspired by \cite{abmaplus}, we selected the following metrics for our comparison.

\emph{Adaptability (A)}  is the average selected video bit-rate per segment in a stream over the minimum of either the average throughput available during the current segment $\overline{C}$ or the maximum available representation $R_{N}$
\begin{equation}
A=\frac{1}{K}\sum_{i=1}^K\frac{{R_i}}{\min(R_{N},\overline{C_i})}.
\end{equation}
This metric may take values above 1, when the algorithm is aggressive, which may lead to un-smoothness. 

Instability consists of the adaptation frequency and, complementary to that, the amplitude of adaptation.
\emph{Adaptation frequency (AF)} is the  number of representation switches over the total number of segments $K$, given by
\begin{equation}
AF=\frac{\sum_{i=1}^{K-1} (1-\delta_{R_{i}R_{i+1}})}{K},
\end{equation}
where $\delta$ is the \emph{Kronecker delta}. 
\emph{Adaptation amplitude (AA)}  is the normalized average distance, in terms of bit-rate, between the representation levels.
\begin{equation}
AA=\frac{1}{K\cdot AF}\frac{\sum_{i=1}^{K-1} |R_{i+1}-R_{i}|}{ R_{max}}.
\end{equation}

When considering un-smoothness we must take into account the re-buffering duration along with the frequency of re-buffering events. \emph{Re-buffering  duration (RD)} is the total duration of re-buffering events  in a stream over the length of the played-out video $L$,
\begin{equation}
RD=\frac{\sum_{i=\omega+1}^K \beta_i\cdot(t_i^{end}-t_i^{start})}{L},
\end{equation}
where $\beta_i=1$ if a re-buffering event occurred during the download of segment $i$ and $\beta=0$ otherwise and $t_i^{end}$ and $t_i^{start}$ are the time of end and the start of the re-buffering event, which occurred during the download of segment $i$, respectively.
\emph{Re-buffering frequency (RF)} is the number of re-buffering events that occurred in a stream over the number of segments $K$
\begin{equation}
RF=\frac{\sum_{i=\omega+1}^K \beta_i}{K}.
\end{equation}

In the figures of Section \ref{resultssection}, these metrics are averaged over the complete set of traces for every profile and we present the standard error with a confidence level of $95\%$.

\section{Results}
\label{resultssection}
In this section we evaluate the performance of each adaptation algorithm based on the metrics introduced in Section \ref{experimentalframeworksection}. The results are not standalone and a combination of the QoE metrics is required for the performance evaluation of the algorithms. The scope of this paper is not to introduce a QoE model but to present the raw results of the selected metrics.
\subsection{Implementation validation}
\begin{figure}[t]
	\includegraphics[width=1\linewidth,clip]{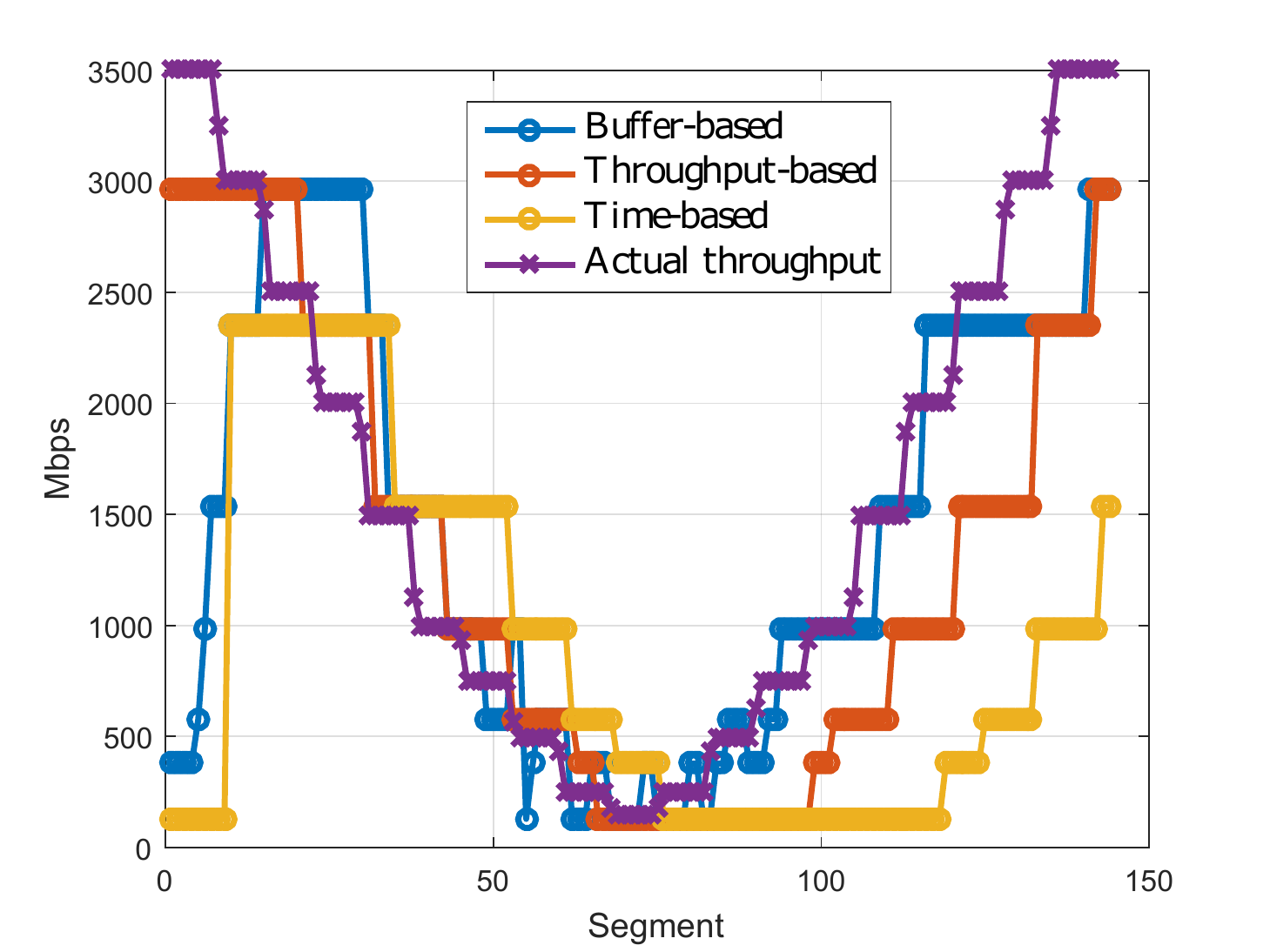}
	\caption{Example of  the representation selection behavior of the three studied classes of algorithms}
	\label{fig:QS}
\end{figure}
 In Fig. \ref{fig:QS} the throughput profile (controlled) that was used to validate our implementations is shown. We can see as a first difference, that the buffer-based adaptation starts with a low representation and gradually increases it while the buffer fills up. On the other hand throughput-based methods estimate the available throughput, through probes, and match it to the respective available representation level. The time-based starts with a low representation until the algorithm has a sufficient amount of probes to estimate the download time appropriately. The second significant difference is that buffer-based  and time-based adaptation may select a representation higher than the available throughput for a short period as potential throughput drops have not, yet, affected the buffer level or registered in the time-probing sample, respectively. On the other hand, the throughput-based algorithms can be more reactive. It is evident that time-based adaptation has a small delay in the adaptation to the current throughput, as the throughput variation is registered in the time-probing sample as an average of the last 50 probes. 
At this moment it is worth mentioning that all studied algorithms were designed using either heuristically or based on  pre-selected parameters. In all our implementations we used the parameterization proposed by the designers, but better results could be achieved if the parameters were fine-tuned.

\subsection{Adaptability}

\begin{figure}[t]
    \includegraphics[ width=1.0\linewidth,clip]{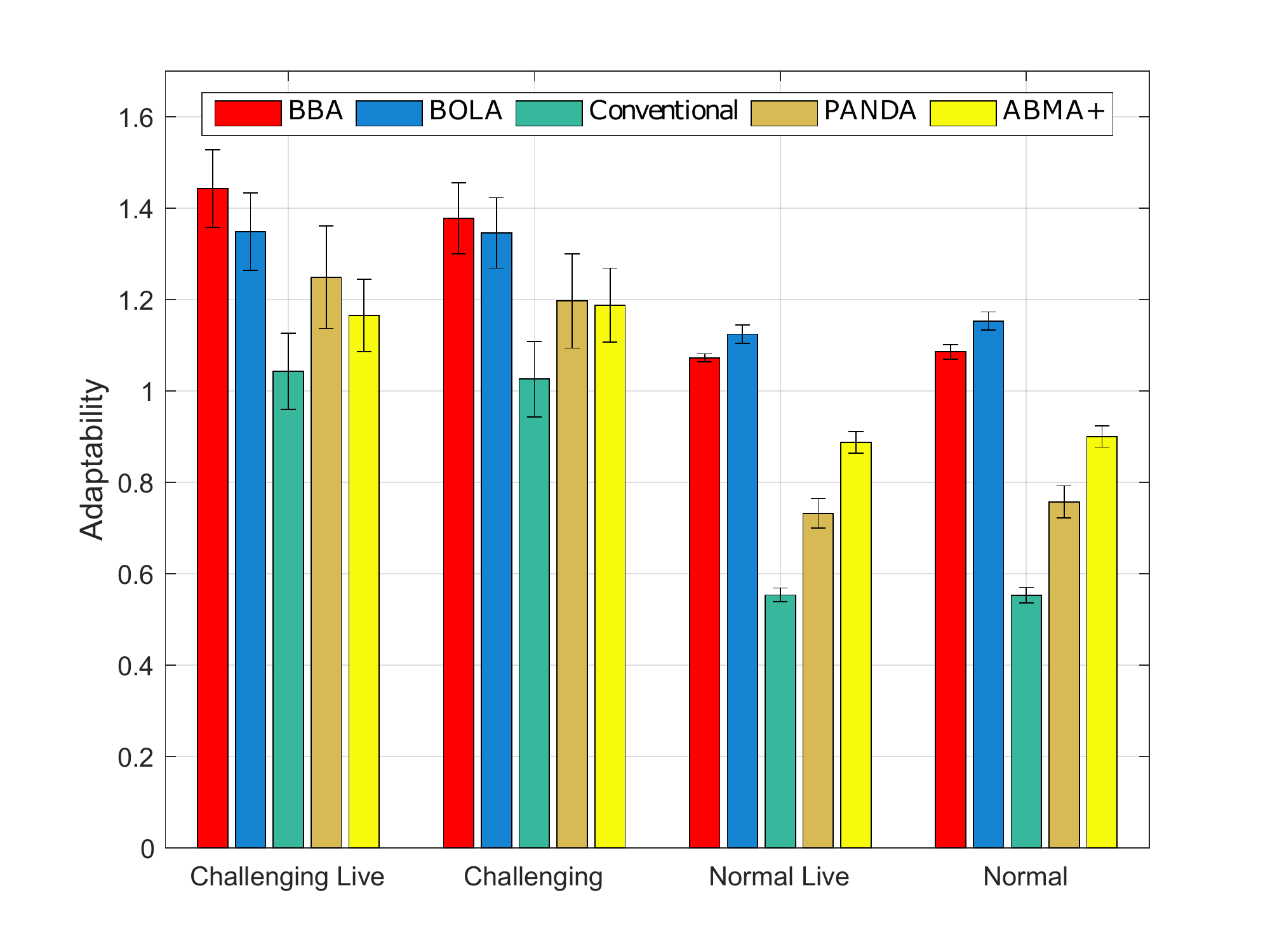} \vspace{-20pt}
     \caption{ Experimental results for adaptability with 95\% confidence intervals }
     \label{fig:RSE}
\end{figure}
   \begin{figure}[t]
   	\includegraphics[ width=1.0\linewidth,clip]{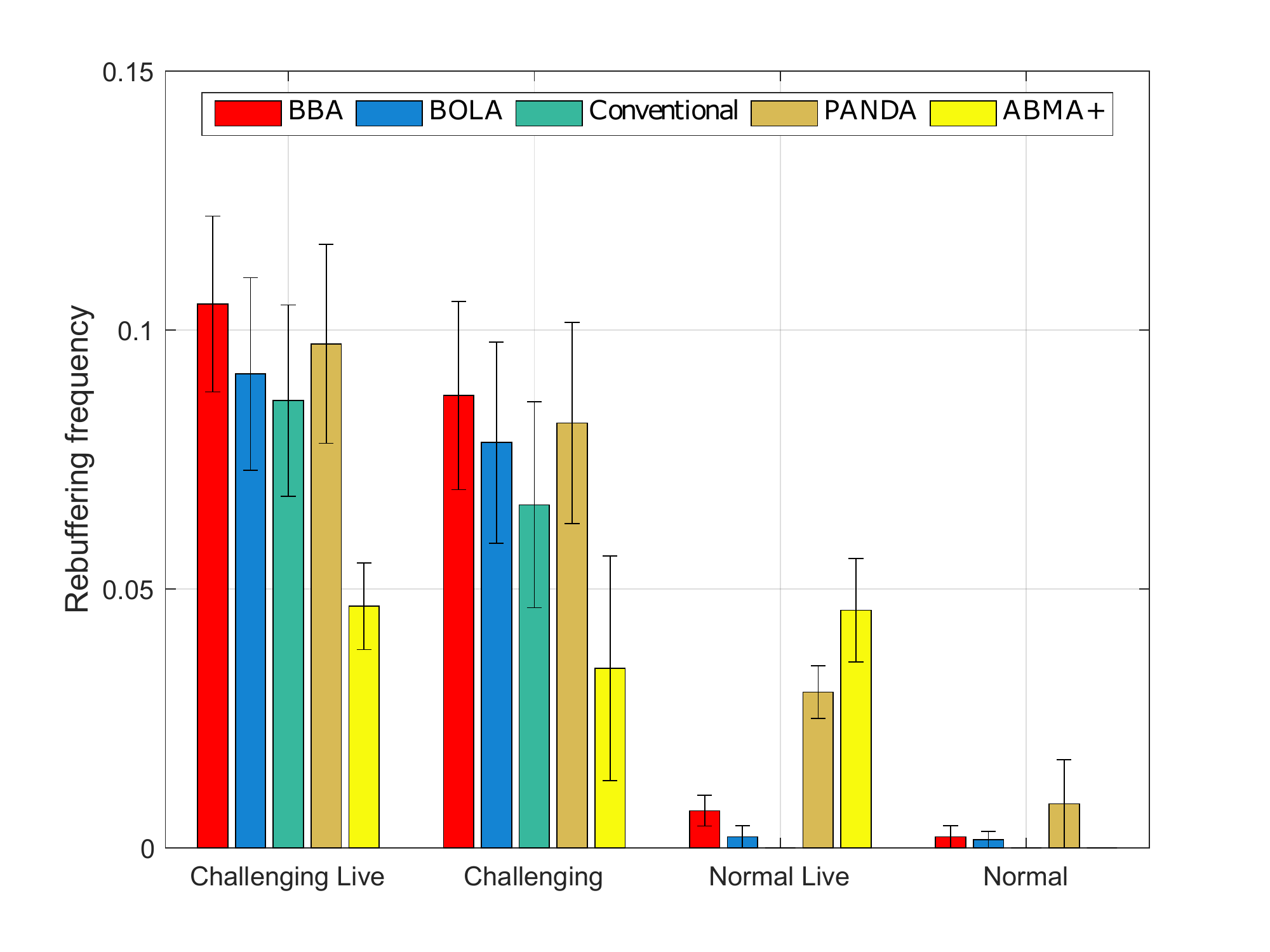} \vspace{-20pt}
   	\caption{ Experimental results for re-buffering frequency with 95\% confidence intervals}
   	\label{fig:/RER}
   \end{figure}
     \begin{figure}[t]
     	\includegraphics[ width=1.0\linewidth,clip]{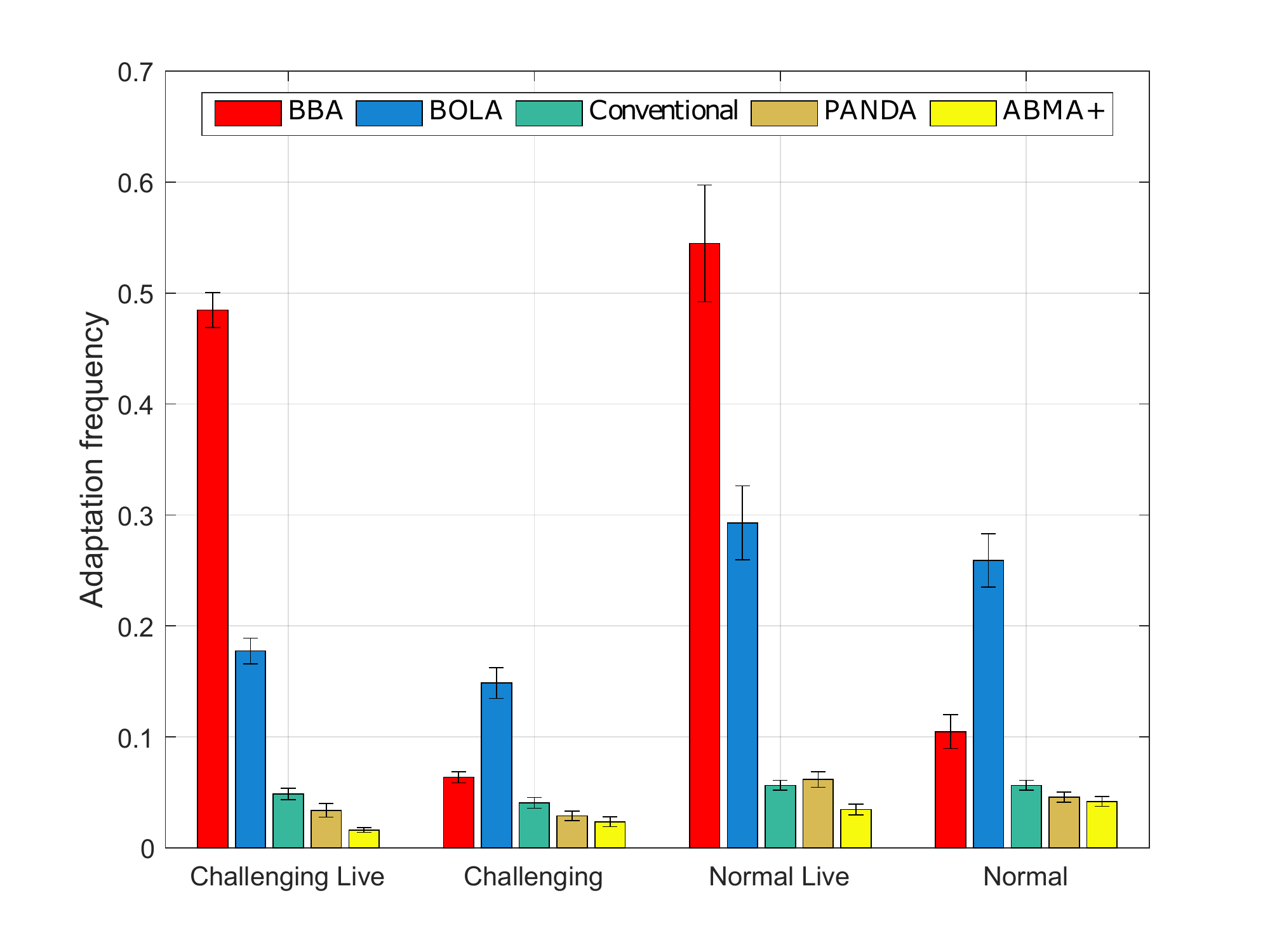} \vspace{-20pt}
     	\caption{ Experimental results for adaptation frequency with 95\% confidence intervals }
     	\label{fig:RSR}
     \end{figure}
     \begin{figure}[t]
     	\includegraphics[ width=1.0\linewidth,clip]{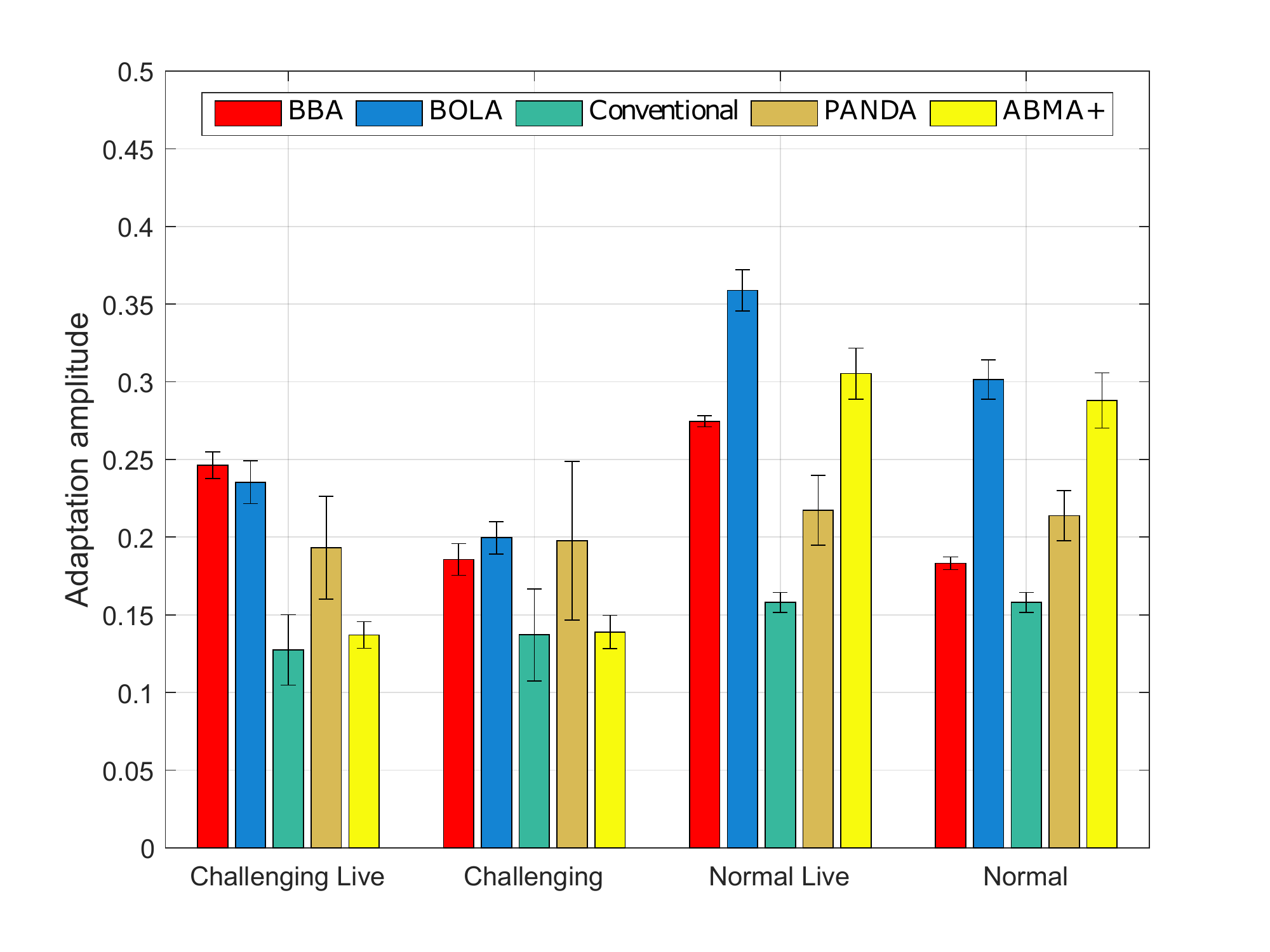} \vspace{-20pt}
     	\caption{ Experimental results for adaptation amplitude with 95\% confidence intervals }
     	\label{fig:/RSA}
     \end{figure}
     \begin{figure}[t]
     	\includegraphics[ width=1.0\linewidth,clip]{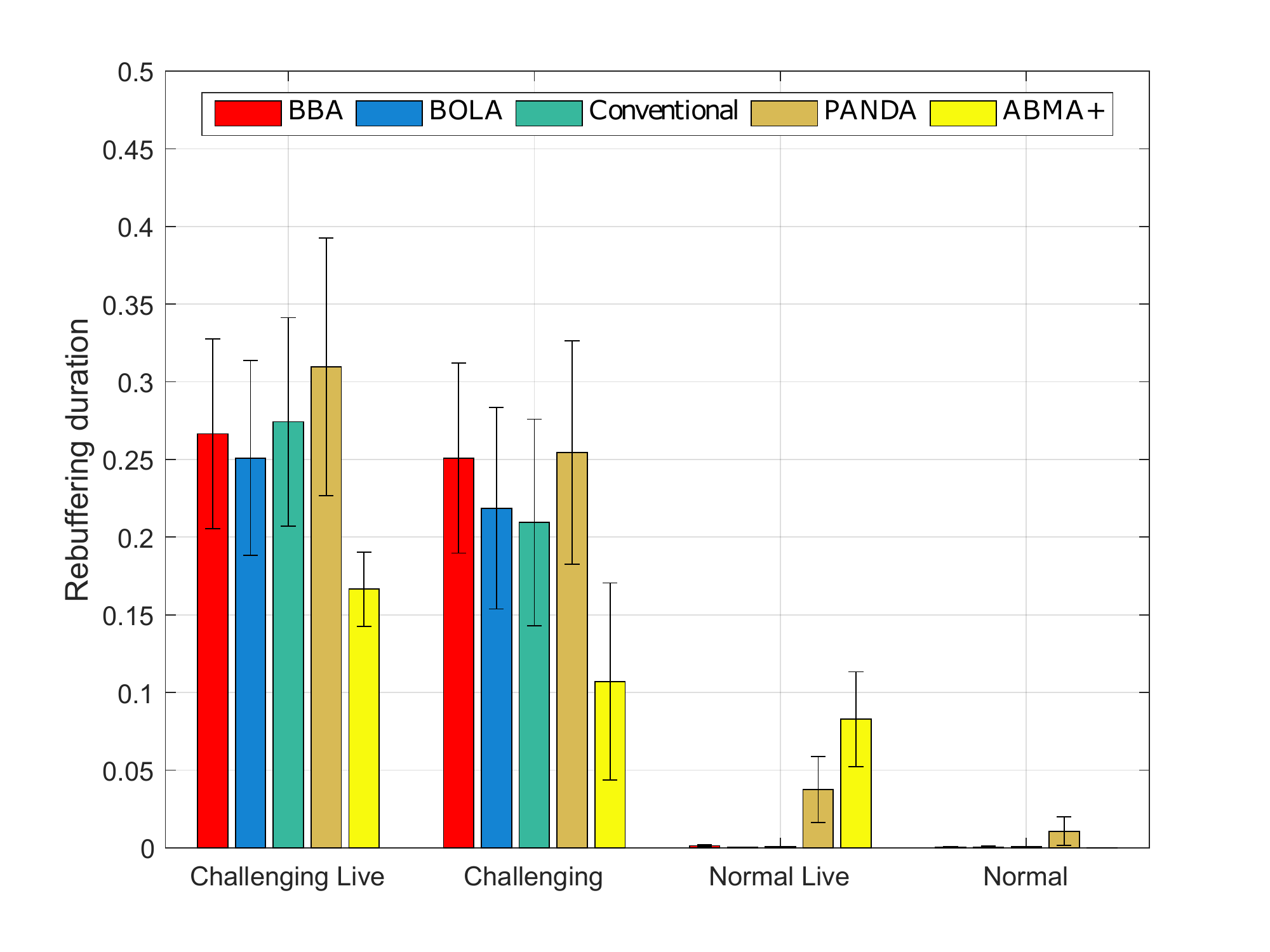} \vspace{-20pt}
     	\caption{ Experimental results for re-buffering duration with 95\% confidence intervals }
     	\label{fig:RED} 
     \end{figure}
     
Fig. \ref{fig:RSE} shows that buffer-based algorithms achieve higher adaptability in normal conditions. They are more successful, by design,  in conserving high representation levels, even higher than the available throughput, as the adaptability becomes larger than 1.  Throughput-based and time-based algorithms show a slightly diminished ability to match the representation to the available average throughput due to the significant throughput variation that characterizes the selected network profiles. We also notice from this figure that the buffer size does not affect significantly the adaptability.

\subsection{Un-smoothness}
As far as un-smoothness is concerned, Fig. \ref{fig:/RER} shows that, as expected, the probability of a re-buffering is slightly higher in the cases of a small buffer (i.e live streaming). A small buffer has limited resilience to throughput variation. Regarding the re-buffering frequencies per class of algorithms, we can note that although the performances are very close, on challenging profiles buffer-based algorithms, along with \emph{PANDA}, are slightly more probable to experience a re-buffering. For normal scenarios we witness smooth streaming from almost all algorithms, due to the absence of long throughput outages of this profile and the design of our simulations (selection of representation levels based on quantiles of normal profile). Nevertheless \emph{AMBA+} shows a slightly increased un-smoothness compared with the rest of the algorithms in the normal profile with a small buffer, due to the fact that the number of probes (50) proposed by the algorithm designers is very high compared to the maximum buffer. Therefore a short throughput drop is not registered in the estimation in appropriate time before the buffer has been depleted. Fig. \ref{fig:RED} shows the duration (amplitude) of the re-buffering events, where we see re-buffering events lasting about 25\% of the video duration. This is expected since the challenging profile  includes underground areas, which may cause network outages, for 1/4 of the trace.

\subsection{Instability}  
Last but not least, instability has a very significant impact on QoE \cite{HAS_QoE}. Fig. \ref{fig:RSR} shows that buffer-based algorithms are about 40\% more probable of making a quality switch when the buffer is small. \emph{BOLA} is optimized to achieve a high efficiency but the stability aspect is not considered in the optimization, since it is addressed with a heuristic in a second phase. \emph{BBA} has a pre-selected constant higher buffer threshold which makes the segment map less agile to throughput variation when the  maximum buffer is small.  On the contrary, throughput-based and time-based algorithms appear to switch quality less often, but with similar adaptation amplitude, which is complementary to adaptation frequency if one wants to draw a conclusion on stability. Fig. \ref{fig:/RSA} shows the average normalized distance between switches. The performance regarding this metric shows a slight advantage in favor of the throughput-based algorithms, in both normal and challenging profiles. 

It is important to mention that no metric should be treated separately and that only the combination of all metrics allows our comparison to be insightful. Overall, our results match those in \cite{BOLA} with the addition that \emph{BOLA} is compared against another buffer-based adaptation for the first time. Moreover, our results can be verified with those in \cite{abmaplus} where \emph{PANDA} and \emph{BBA} are compared against \emph{ABMA+}. In Table \ref{reco} we have gathered the best performing classes of algorithms, per QoE element. This table can serve as insight to the selection of the most appropriate algorithmic class, depending on the application parameters (live, VOD, etc.) and the commonly experienced network conditions.
\begin{table}[t]
\centering
\caption{Results per QoE metric per class} \vspace{-5pt}
\label{reco}
\resizebox{\linewidth}{!}{\begin{tabular}{|c|c|c|c|c|c|c|}
\hline
\textbf{Buffer Size} & \textbf{\begin{tabular}[c]{@{}c@{}}Network\\  condition\end{tabular}} & \textbf{\begin{tabular}[c]{@{}c@{}}Adaptability\end{tabular}} & \textbf{\begin{tabular}[c]{@{}c@{}}\\Re-buffering\\ frequency\end{tabular} } & \textbf{\begin{tabular}[c]{@{}c@{}}Re-buffering \\duration\end{tabular}} & \textbf{\begin{tabular}[c]{@{}c@{}}Adaptation\\ frequency\end{tabular}} & \textbf{\begin{tabular}[c]{@{}c@{}} Adaptation\\ amplitude\end{tabular}} \\ \hline
Small    (16 s)            & Normal       &    Buffer    &   Time  & Time  & Time&  Throughput  \\ \hline
Small     (16 s)           & Challenging  &    Buffer    &   Time        & Time  &Time&Throughput\\ \hline
Large           (92 s)     & Normal       &    Buffer    &   Buffer  & Buffer  &Time&Throughput \\ \hline
Large          (92 s)      & Challenging  &    Buffer    &   Buffer        &  Buffer      & Time&Throughput\\ \hline
\end{tabular}} 
\end{table}


\section{conclusion}
\label{conclusionssection}
In this study, we evaluated the performance of five state-of-the-art adaptive streaming algorithms and made a per class comparison, based on network traces for two different throughput profiles. Additionally, we evaluated the maximum buffer occupancy factor to see how each strategy behaves for smaller and larger buffers, as different services may target different buffer levels. Our conclusion is that buffer-based approaches outperform any other class of algorithms in terms of adaptability, yet they may lack in stability, especially for small buffers, common in live streaming services. 

This work provides first guidelines to designers and operators of HAS algorithms for the right algorithmic approach according to expected network conditions and service requirements. Designing \emph{robust} HAS algorithms for high QoE under changing conditions and requirements, without relying on pre-selected designer-specific parameters or heuristic design, is still a major challenge for research.

\bibliographystyle{ACM-Reference-Format}
\bibliography{ConfAbrv,IEEEabrv,JRNLabrv,theo} 

\end{document}